\documentclass{appolb}
\usepackage{graphicx}

\usepackage[square,numbers,sort&compress]{natbib}

\usepackage{mathtools} 
\usepackage{mleftright} \mleftright 
\usepackage[dvipsnames]{xcolor}
\usepackage{xspace}
\usepackage{bm} 
\usepackage{booktabs} 
\usepackage{hyperref}
\hypersetup{
  colorlinks   = true,
  urlcolor     = blue, 
  linkcolor    = blue, 
  citecolor   = blue 
}
\usepackage[nameinlink]{cleveref} 

\newcommand{\leftd}{\left.\kern-\nulldelimiterspace}
\newcommand{\rightd}{\right.\kern-\nulldelimiterspace}
\newcommand{\rcite}[1]{Ref.~\cite{#1}}
\newcommand{\rscite}[1]{Refs.~\cite{#1}}
\newcommand{\unit}[1]{\ \text{#1}} 
\newcommand{\dd}{\mathrm{d}} 
\newcommand{\vnabla}{\bm{\nabla}}
\newcommand{\vR}{\mathbf{R}}
\newcommand{\vx}{\mathbf{x}}
\newcommand{\vJ}{\mathbf{J}}
\newcommand{\Ekin}{E_\text{kin}}
\newcommand{\EH}{E_\text{H}^\chi}
\newcommand{\EF}{E_\text{F}^\chi}
\newcommand{\ESk}{E_\text{Skyrme}}
\newcommand{\ECou}{E_\text{Coulomb}}
\newcommand{\Epair}{E_\text{pair}}
\newcommand{\SatDens}{\rho_c} 
\newcommand{\SatEn}{E_\text{sat}} 
\newcommand{\Incomp}{K} 
\newcommand{\SymEn}{a_\text{sym}} 
\newcommand{\SlopePar}{L_\text{sym}} 
\newcommand{\ScEffMassInv}{M_\text{s}^{\ast -1}} 
\newcommand{\elm}[2]{$^{#2}$#1} 
\newcommand{\NNLO}{N$^2$LO\xspace}
\newcommand{\NLOD}{NLO$\Delta$\xspace}
\newcommand{\NNLOD}{N$^2$LO$\Delta$\xspace}
\newcommand{\NLODThreeN}{NLO$\Delta$+3N\xspace}
\newcommand{\NNLOThreeN}{N$^2$LO+3N\xspace}
\newcommand{\NNLODThreeN}{N$^2$LO$\Delta$+3N\xspace}

\newcommand{\nochi}{no chiral\xspace}
\newcommand{\minchi}{min.\ chiral\xspace}

\newcommand{\PenaltyFunc}{loss function\xspace}
\newcommand{\cov}{\mathbf{V}}
\newcommand{\vsigma}{\bm{\sigma}}

\crefname{table}{Table}{Tables}
\crefname{figure}{Fig.}{Figs.}
\Crefname{figure}{Figure}{Figures}
\crefname{equation}{Eq.}{Eqs.}
\Crefname{equation}{Equation}{Equations}
\crefname{section}{Sec.}{Secs.}
\Crefname{section}{Section}{Sections}

\begin{document}
\title{Towards nuclear energy density functionals from first principles%
\thanks{Presented at the 57th Zakopane Conference on Nuclear Physics, {\it Extremes of the Nuclear Landscape}, Zakopane, Poland, 25 August–1 September, 2024.}%
}
\author{Lars Zurek
\address{CEA, DAM, DIF, 91297 Arpajon, France \\
and \\
Université Paris-Saclay, CEA, LMCE, 91680 Bruyères-le-Châtel, France}
}
\maketitle
\begin{abstract}
We discuss the GUDE functionals which consist of pion exchanges derived from chiral effective field theory and a Skyrme-like part.
Certain pion terms lead to significant improvements in the description of ground-state energies, indicating they might be useful ingredients for true ab initio energy density functionals.
In addition, we present estimates of the statistical parameter uncertainties of the GUDE functionals. 
\end{abstract}
  
\section{Introduction}

Nuclear energy density functionals (EDFs)~\cite{Bend03RMP,Schu19EDFBook} constitute an approach to the nuclear many-body problem that occupies a middle ground between more phenomenological approaches, such as the semi-empirical mass formula, and more microscopic ones, such as the so-called ab initio approaches.
This holds both in terms of computational complexity and in terms of resolved degrees of freedom.
As EDFs constitute a mean-field method, they possess mild computational scaling with mass number.
Hence, they can be employed for computing observables throughout the entire nuclear chart (with exception of the very lightest nuclei for which a mean-field approximation is poorly justified). 
In addition, EDFs are at the moment generally able to achieve more accurate results than ab initio methods.

However, the successful description of experiment is, at least in part, a consequence of the phenomenological construction of the employed functional forms, in which parameters are obtained by fitting to experimental results over a wide mass range.
The empirical nature of EDFs makes extrapolations outside the fitting regions, as they are needed e.g.\ for astrophysical applications, potentially uncontrolled~\cite{Mump16IndiRPro,McDo15UncQuaEDF}.
For similar reasons, it is difficult to assess model uncertainties, which would be needed for a full understanding of the predictive capabilities of different functionals.  
Additionally, it is believed that further improvements of EDF accuracy will hardly be obtained just from more and more sophisticated parameter optimization protocols and new forms of the employed functionals are called for~\cite{Kort08SkyrmeSP}.
This proceedings article deals with one approach for obtaining new EDF forms by making connections to chiral effective field theory (EFT).

\section{Quest for energy density functionals from first principles}

Ab initio methods promise to overcome several problems of nuclear EDFs.
This is because they are based on systematically improvable interactions obtained from chiral EFT, which is rooted in quantum chromodynamics.
Their systematical construction allows for a natural way of estimating interaction uncertainties.
Similarly, the many-body problem is solved with systematically improvable correlation expansion methods like many-body perturbation theory or the coupled cluster method (CC).
These many-body methods treat correlations of particle-hole type, which in the EDF approach are generally assumed to be implicitly accounted for by the effective nature of the employed functionals, explicitly.
Therefore, such ab initio calculations are computationally much more expensive than mean-field calculations.
Hence, it is of interest to meaningfully connect nuclear EDFs and the ab initio approach (see, for instance, the parallel drawn in \rcite{Dugu22PGCMEDF}).

Different strategies have been pursued to this end such as fitting the form of EDF volume terms to equations of state obtained from ab initio calculations of infinite nuclear matter (INM)~\cite{Bulg18SeaLL1}.
In addition, fitting to other ab initio pseudodata has also been considered, e.g., to energies of neutron drops~\cite{Bonn18EDFNDrop}.
There are also ideas for obtaining EDFs fully from first principles in the form of effective actions, for instance, in path-integral approaches.
While this constitutes the most fundamental of the research lines mentioned here, there are still many steps to be taken until a fully microscopic nuclear EDF will be constructed~\cite{Furn20EDFasEFT}.

In this proceedings article, we report on findings obtained following a different route, which was first suggested in \rscite{Drut10PPNP,Gebr10DME}.
We concentrate here on the results obtained in the newest work \cite{Zurek24GUDE} (see references therein for previous steps), in which we constructed the GUDE\footnote{Germany-USA Density-matrix expansion Energy density functionals} family of functionals, which can be regarded as semi-phenomenological or hybrid functionals.
They are obtained by complementing conventional Skyrme-type contact terms with terms arising from pion exchanges as described by chiral EFT at the Hartree-Fock level.

This strategy can be motivated in two ways.
First, we note that ab initio calculations in heavy systems typically proceed by solving a Hamiltonian obtained from chiral EFT on the mean-field level. 
Then, correlations are build on top of this solution via a correlation expansion method.
Instead of explicitly generating correlations by the many-body method, our approach can be viewed as an adjustment of the short-range part of the potential reflecting the fact that dominant bulk correlations in nuclei appear to be of short-range nature.
This is similar in spirit to attempts of encapsulating the effect of triples correlations in the coupled-cluster approach into CCSD by adjusting the three-nucleon contact interaction~\cite{Sun22RenormCC}.
Second, conventional Skyrme EDFs correspond largely to effective contact interactions solved at the mean-field level.
When increasing the resolution of the description of the considered systems, the first additional degree of freedom that is resolved are exchanges of the lightest mesons, the pions, as described by chiral EFT.
Hence, including them explicitly can be expected to lead to a more accurate description of nuclei.

\section{Construction and optimization of GUDE functionals}

We now proceed with outlining the form of functionals in the GUDE family as well as the parameter optimization.
Here, we restrict ourselves to the most important points and refer the reader to \rcite{Zurek24GUDE} for details regarding implementation and optimization protocol.

The functionals, which are solved at the Hartree-Fock-Bogoliubov level, consist of six parts, 
\begin{equation}
    E = \EH + \EF + \ESk + \ECou + \Epair + \Ekin \,,
\end{equation}
which can be separated into a conventional Skyrme-like structure (the latter four terms) and long-range contributions from pion exchanges (the first two).
The pion exchanges are taken from chiral EFT and are considered at the Hartree-Fock level.
The different functionals within the GUDE family differ by which pion-exchange terms are included.
We consider terms at different orders in the chiral expansion up to next-to-next-to-leading order (\NNLO).
At each order, we construct functionals with and without the explicit inclusion of intermediate $\Delta$ isobars as well as with and without the inclusion of 3N forces.
In total, this yields eight GUDE variants: LO, NLO, \NNLO, \NNLOThreeN, \NLOD, \NLODThreeN, \NNLOD, \NNLODThreeN.
Note that the pion exchanges go beyond the simple Yukawa-type one-pion exchange which appears at LO in the chiral expansion.

For the Hartree terms $\EH$, the potentials are approximated by sums of Gaussians which allows a simpler implementation into existing EDF codes since they are often already capable of treating such terms because they occur in the widely used family of Gogny functionals.
Fock contributions $\EF$ are included in quasi-local form.
This is achieved by using a density-matrix expansion (DME) to approximate the one-body density matrix in terms of quasi-local densities and leads for the NN forces schematically to
\begin{align}\label{eq:NNFock}
    E_\text{F,NN}^\chi = \sum_{t=0,1} &\int \! \dd\vR \,
    \left\{ g_t^{\rho\rho}(\rho_0) \rho_t^2 
    + g_t^{\rho\tau}(\rho_0) \rho_t \tau_t 
    \vphantom{g_t^{JJ,3}} 
    + g_t^{\rho\Delta\rho}(\rho_0) \rho_t \Delta\rho_t
    \rightd\notag \\
    &\null 
    + \leftd g_t^{JJ,2}(\rho_0) J_{t,ab} J_{t,ab}
    + g_t^{JJ,1}(\rho_0) \left[ J_{t,aa} J_{t,bb}
    + J_{t,ab} J_{t,ba} \right] \right\} \,.
\end{align}
where, unlike for contact interactions, all the $g$ coefficients are functions of the isoscalar density $\rho_0$.
In \cref{eq:NNFock} we suppressed the dependence of the various (quasi-)local densities on $\vR$, and $t=0$ ($t=1$) labels isoscalar (isovector) quantities.
The DME is in particular a suitable strategy to make the inclusion of 3N interactions computationally feasible.
In this case, one obtains a similar (but longer) equation to \cref{eq:NNFock}, with density trilinears instead of bilinears.

In addition to the different chiral terms, we include for all GUDE variants a Skyrme part,
\begin{align}
    \ESk = \sum_{t=0,1} &\int \! \dd\vR \,
    \bigl[ \left(C_{t0}^{\rho\rho} + C_{tD}^{\rho\rho} \rho_0^\gamma\right) \rho_t^2 
    + C_t^{\rho\tau} \rho_t \tau_t 
    + C_t^{\rho\Delta\rho} \rho_t \Delta\rho_t
        \notag \\
    &\null 
    + C_t^{\rho\nabla J} \rho_t \vnabla \cdot \vJ_t
    + C_t^{JJ} J_{t,ab} J_{t,ab}
        \bigr] \,,
\end{align}
in addition to a Coulomb contribution $\ECou$ and the kinetic energy $\Ekin$.
Nuclear superfluidity is described by a contact mixed-pairing term $\Epair$.

For every GUDE variant, we fit the parameters that appear in $\ESk$ after adding the pion exchanges.
Note that we take the pion exchanges as they are from chiral EFT, where the low-energy constants are fitted to pion-nucleon scattering data~\cite{Kreb07Deltas}.
Therefore, they are free of parameters to be adjusted in the many-body sector.
We fix the poorly constrained isovector effective mass at its SLy4 value, which leaves $n_x=14$ Skyrme parameters to be optimized for each GUDE variant.

The parameters $\vx$ are optimized by minimizing a \PenaltyFunc
\begin{equation}
    \chi^2(\vx) = \Vert\vR(\vx)\Vert^2
\end{equation}
with $\vR(\vx)$ being an $n_d$-dimensional vector, whose components
\begin{equation}
    R_i(\vx) = \frac{s_i(\vx)-d_i}{w_i}
\end{equation}
are weighted residuals of the EDF predictions $s_i(\vx)$ and experimental data $d_i$.
It consists of $n_d=121$ data points for 81 even-even nuclei ranging from \elm{Ca}{40} to \elm{Hs}{264}, namely binding energies, charge radii, odd-even mass staggerings, and fission isomer excitation energies.
The data set is similar to the one used in \rscite{Kort14UNEDF2, Nava18DMEEDF}.
The weights $w_i$ are taken from a Bayesian calibration of the UNEDF1 functional~\cite{Schu20EDFCalibr}.
To ensure that the GUDE functionals provide a reasonable description of infinite nuclear matter, we perform the optimizations with bounds on INM parameters, which are constrained to a physically plausible region.

Note that in addition to the eight GUDE variants including pion exchanges up to different orders, we also construct an EDF without any chiral terms, but following the same optimization protocol.
This Skyrme functional, which we call ``\nochi'', serves as our reference functional, to which we compare the other GUDE variants.
A tenth GUDE variant is obtained by including only the chiral contributions which we find to significantly influence the performance of the GUDE functionals, see \cref{sec:results}. 
We refer to this functional as the ``\minchi'' variant.

\section{GUDE parametrizations and uncertainties}

In \cref{tab:param} we provide the obtained EDF parameters for the ``\nochi'' and ``\minchi'' functionals, where EDF volume parameters have been expressed in terms of INM properties.
For the other parametrizations, we refer the reader to \rcite{Zurek24GUDE}.

We want to focus here on the calculation of statistical uncertainty estimates of the different parameters.
We obtain these estimates by computing a first-order approximation to the covariance matrix $\cov$ based on the \PenaltyFunc with its minimum at $\vx$:
\begin{equation}
    \cov \approx \frac{\chi^2(\vx)}{n_d-n_x} \left( \vJ(\vx)^T \vJ(\vx) \right)^{-1} \,.
    \label{eq:cov}
\end{equation}
The Jacobian $\vJ(\vx)$ is obtained from
\begin{equation}
    \vJ(\vx)^T = \vnabla \otimes \vR(\vx) \,,
\end{equation}
and parameter standard deviations $\vsigma$ are then given by the square roots of the diagonal elements of $\cov$.
\Cref{eq:cov} constitutes a well-known method to approximate covariance matrices in the field of non-linear regression~\cite{Donaldson87NonlinearLS,Seber2003NonlinearRegressionBook} and has been employed for instance in the context of optimizing Fayans~\cite{Reinhard24Fayans} and the UNEDF~\cite{Kort10edf,Kort14UNEDF2} functionals.
This notion of a covariance matrix assumes that \cref{eq:cov} is evaluated at a point $\vx$ where $\chi^2(\vx)$ has a true minimum (in the sense that it can be locally approximated by a quadratic function in the direction of every component $\vx_j$).
However, this is not true for any GUDE variant:
for all variants, the optimization yields a minimum where at least one INM parameter ends up at one of its bounds. 
To be able to apply \cref{eq:cov} anyway, we therefore restrict our analysis to the subspace of parameters that do not end up actively constrained by the bounds and evaluate the Jacobian only in the directions corresponding to that subspace.
See \rcite{Kort10edf} for a related discussion.

Numerically, we compute the Jacobian by means of central differences using the points $\{ \vx \pm \eta_j e_j \}$, where $e_j$ are unit vectors in the directions of the different parameters and the step widths $\eta_j$ should be small.
We choose $\eta_j$ as $10^{-5}$ of the scaling intervals of the different parameters given in \rcite{Wild15POUNDEDF}.\footnote{For the $C_t^{JJ}$ parameters, which were not included in \rcite{Wild15POUNDEDF}, we use [-100, 100] MeV~fm$^5$ as scaling intervals.}
For one functional, we check explicitly that this choice of $\eta_j$ is small enough (but not too small to be dominated by numerical accuracy) by comparing against results obtained with a ten times larger step width.
The resulting standard deviations change by less than 10\%, which hence can be interpreted as an uncertainty of the standard deviations reported here.

In \cref{tab:param} we provide the parameter standard deviations $\vsigma$ that result from the procedure explained above for the ``\nochi'' and ``\minchi'' functionals.
In the ``\minchi'' optimization, $\SlopePar$ attained a value at the bound and is thus excluded from the analysis.
Therefore, no value is given for its standard deviation.\footnote{Similar considerations apply to $\SatEn$ and $\Incomp$ in the ``\nochi'' case.}
Due to correlations between the parameters, this also has consequences for other estimated standard deviations.
In particular, the reported standard deviation of $\SymEn$ is expected to be strongly underestimated as $\SlopePar$ and $\SymEn$ are typically strongly correlated.
The remaining isovector parameters show large standard deviations, signalling they are poorly constrained by the used fit data.
We observe this also by comparing minima obtained with optimization runs starting from different initial guesses.
This is typical for current EDFs~\cite{McDo15UncQuaEDF,Naza14SymEnDFT}.

\begin{table}[htb]
    \caption{\label{tab:param}
    Parameters of two GUDE variants and associated standard deviations given in form of statistical uncertainties $\vx (\vsigma)$.
}
\centerline{%
\begin{tabular}{lrr}
\toprule
 & \nochi & \minchi \\ \midrule 
$\SatDens \unit{(fm$^{-3}$)}$ & 0.1546(4) & 0.1583(9) \\ 
$\SatEn \unit{(MeV)}$ & $-15.8$ & $-15.830(17)$ \\ 
$\Incomp \unit{(MeV)}$ & 260 & 224(7) \\ 
$\ScEffMassInv$ & 0.979(3) & 0.917(3) \\  
$\SymEn \unit{(MeV)}$ & 29.9(8) & 28.6(3) \\ 
$\SlopePar \unit{(MeV)}$ & 41(16) & 30 \\ 
$C_0^{\rho \Delta\rho} \unit{(MeV fm$^5$)}$ & $-41.4(8)$ & 22.5(1.0) \\
$C_1^{\rho \Delta\rho} \unit{(MeV fm$^5$)}$  & $-6(26)$ & $-39(19)$ \\
$C_0^{\rho \nabla J} \unit{(MeV fm$^5$)}$ & $-62(6)$ & $-61(5)$ \\
$C_1^{\rho \nabla J} \unit{(MeV fm$^5$)}$ & 11(13) & 3(15) \\
$C_0^{JJ} \unit{(MeV fm$^5$)}$ & $-43(17)$ & $-39(15)$ \\
$C_1^{JJ} \unit{(MeV fm$^5$)}$ & $-30(19)$ & $-4(22)$ \\
$V_0^\text{n} \unit{(MeV fm$^3$)}$ & $-218.4(1.4)$ & $-206.5(1.2)$ \\
$V_0^\text{p} \unit{(MeV fm$^3$)}$ & $-259.9(2.4)$ & $-249.4(2.0)$  \\
\bottomrule
\end{tabular} 
}
\end{table}

The standard deviations can be interpreted as statistical uncertainties that give a range of reasonable parameter values within the given model and fitting protocol.
On top of the limiting treatment of actively bound parameters explained above, these statistical uncertainties cannot be considered full uncertainties of the GUDE functionals as systematic errors are not explicitly taken into account.
Such model uncertainties arise from the EDF structure being incomplete or wrong~\cite{Doba14error} and are particularly hard to model for nuclear EDFs, for which a unifying construction principle is not known. 
However, their existence can be clearly inferred, for instance from the systematic trends of the mass residuals around shell closures, see \cref{fig:residuals} below.

\begin{figure}[htb]
\centerline{%
\includegraphics[width=12.5cm]{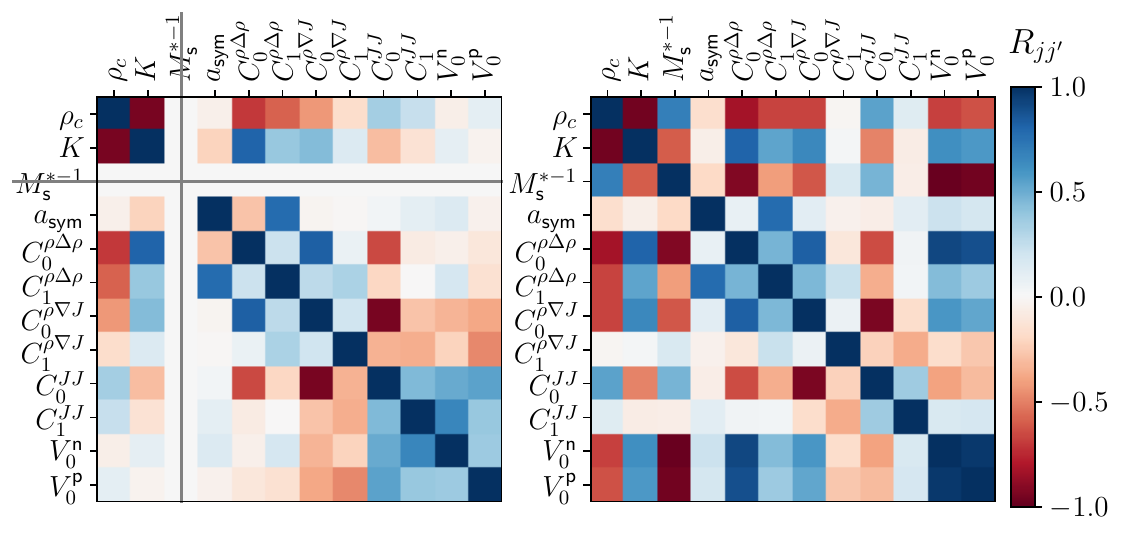}
}
\caption{Parameter correlation coefficients of the \NNLOThreeN GUDE functional. In the left panel the coefficients are calculated in a parameter subspace without $\ScEffMassInv$, $\SatEn$, and $\SlopePar$, while in the right panel only the latter two (which end up at their bounds in the optimization) are excluded.
}
\label{fig:correlations}
\end{figure}

Often, correlation coefficients 
\begin{equation}
    R_{jj'} = \frac{V_{jj'}}{\sigma_j \sigma_j'}
\end{equation}
(or coefficients of determination $R_{jj'}^2$) are used to analyze correlations between different parameters.
To compare correlation coefficients of different GUDE variants in the given framework, we need to compute the covariance matrices in the same parameter subspace for all functionals of interest.
We find that some correlation coefficients depend strongly on which subspace is used to compute them. 
For example, correlation coefficients of the \NNLOThreeN functional change strongly once the isoscalar effective mass $M_\text{s}^{\ast}$ is included in the analysis, see \cref{fig:correlations} (while for the ``\minchi'' variant barely any change occurs).
As for all functionals at least one parameter ends up at its bound in the optimizations, we are not able to calculate correlation coefficients in the full parameter space.
Thus, we abstain here from a detailed analysis of the correlation coefficients.
In the future, an investigation of how the inclusion of pion-exchange terms changes the parameter landscape would certainty be of interest.

\section{Results and analysis}\label{sec:results}

We now turn to a brief investigation of the different GUDE variants obtained from the optimization to address which pion contributions could be useful ingredients for true ab initio EDFs.
A more detailed examination can be found in \rcite{Zurek24GUDE}. 

We find that many of the functionals behave similarly.
Therefore, we group them into three classes:
we refer to the reference ``\nochi'' functional as class~0, to the LO and NLO functionals collectively as class~1, and to the remaining functionals as class~2.
While the value of the $\chi^2$ at the minimum shows no improvement for class~1 over class~0, the opposite is the case when switching to class~2.
All functionals in this class have $\chi^2 \approx 90$, to be compared with the clearly worse value of about 120 for the ``\nochi'' variant, see \cref{tab:results}.

\begin{table}[htb]
    \caption{\label{tab:results}
Performance of the different GUDE classes. 
We list the value of the $\chi^2$ at the optimum and the binding energy root-mean-square deviations when comparing to experiment. 
The latter are calculated from all measured even-even nuclei with $Z \geq 8$ included in the 2020 AME.
The given ranges comprise all functionals within a given class.
}
\centerline{%
\begin{tabular}{lrrr}
\toprule
 & class 0 & class 1 & class 2 \\ \midrule 
$\chi^2$ & 122 & 145 & 86 - 91 \\ 
Binding energy RMSD (MeV) & 2.11 & 2.09 - 2.13 & 1.41 - 1.56 \\
\bottomrule
\end{tabular} 
}
\end{table}

Differences between the classes can also be observed in terms of EDF parameters, e.g., see \cref{tab:param}.
Incompressibility, isoscalar effective mass and therewith correlated pairing strengths, as well as the isovector INM parameters clearly change when going from class~0 or~1 to class~2.
Compared to that, the variations within class~2 are much smaller, which affirms our grouping of functionals.

For a global comparison with experiment, we compute ground states of all even-even nuclei in the nuclear chart with $Z \geq 8$ for which measured binding energies are included in the 2020 AME~\cite{Wang21AME20}.
The root-mean square deviations between the different GUDE variants and experimental binding energies are given in \cref{tab:results}.
A significant reduction by about 30\% is obtained when switching from class~0 or~1 to class~2.
In addition, the mean deviation is almost halved, indicating that the energies are less biased (towards underbinding) for class~2.

\begin{figure}[htb]
\centerline{%
\includegraphics[width=10cm]{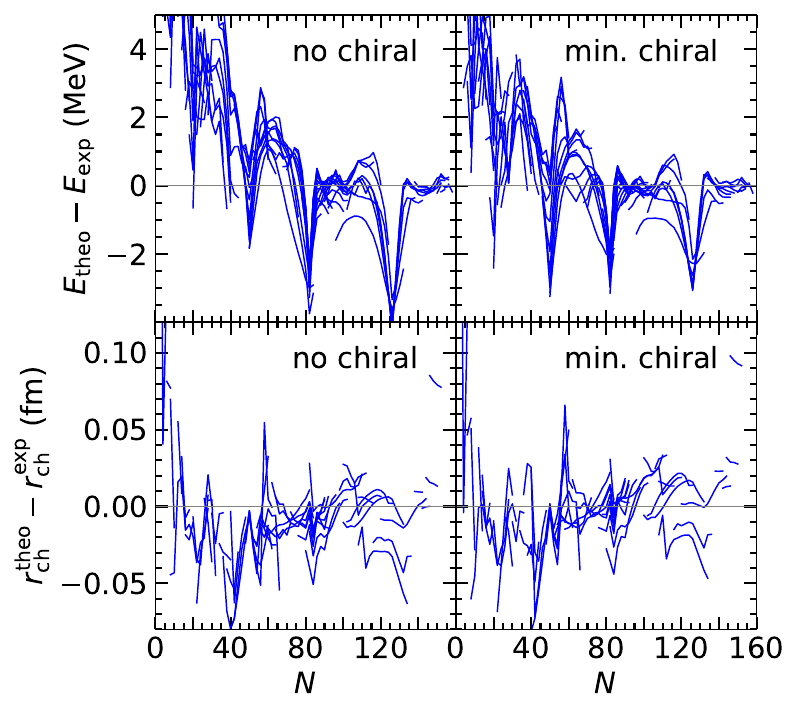}
}
\caption{Differences of ground-state energies (upper panels) and charge radii (lower panels) for even-even nuclei between values obtained
with the ``\nochi'' and ``\minchi'' GUDE variants and experiment.
Figure adjusted from \rcite{Zurek24GUDE}.
}
\label{fig:residuals}
\end{figure}

In \cref{fig:residuals} we show ground-state energy
residuals for the ``\nochi'' and ``\minchi'' functionals.
It becomes apparent that the main binding-energy improvements observed for class~2 occur, on the one hand, around the $N = 82$ and $N = 126$ shell closures and, on the other hand, for light nuclei.
The quality of the description of charge radii (compared to data from \rcite{Ange13rch}), also plotted in \cref{fig:residuals}, is largely insensitive to the GUDE variant considered.
Only a slightly better description for $N \approx 40$ to 100 is observed for class~2.

As a simple validation check that basic features of nuclear shell structures are properly reproduced, we compute single-particle levels using blocking calculations in a few nuclei. 
The obtained shell gaps are very similar for all GUDE variants and are in good agreement with shell gaps extracted from experiment.

In terms of analyzing the effects of the chiral contributions included in the GUDE family, we restrict ourselves here to two points:
First, 3N pion exchanges seem to essentially have no effect in the present construction.
This is probably because the optimization of the parameters in the Skyrme part of the EDFs effectively incorporates the effect of such 3N terms even in the functionals which do not include these 3N terms explicitly.
In particular the density-dependent Skyrme contact term takes care of this, in agreement with the original reason to include such a term in nuclear EDFs.

Second, we find that including just two chiral contributions on top of the Skyrme structure is enough (after refitting the Skyrme parameters) to obtain the improvements observed for the class~2 functionals.
These are the isoscalar NN pion-exchange Hartree contribution entering at \NNLO and the isoscalar NN pion-exchange Fock contribution at LO.
The ``\minchi'' GUDE variant contains precisely only these two chiral contributions.
We reason in \rcite{Zurek24GUDE} why including these chiral terms yields the obtained improvements and why the other chiral terms do not need to be included.

The observed behavior of the functionals in terms of chiral orders might seem to be in contradiction to the chiral EFT power counting, according to which the importance of additional terms is reduced with every higher order included.
We believe this is because we only consider mean-field contributions, whereas the chiral power counting is formulated for the full potential, and because the structure of the contact interactions in the GUDE functionals (i.e., the Skyrme part) does not change with increasing order.

\section{Conclusions}

In this proceedings article, we reported on the GUDE EDFs which we constructed by adding chiral long-range pion-exchange terms to a Skyrme-like structure.
Additionally, we presented here statistical uncertainty estimated for the obtained parametrizations.

When including chiral contributions beyond NLO, we find a significant improvement of the reproduction of experimental binding energies (while being constrained to deliver reasonable infinite matter properties) compared to a reference Skyrme EDF optimized according to the same protocol.
Since the chiral contributions are added in a parameter-free manner, the origin of the improvements has to lie in the structure of the considered chiral contributions itself.
This suggests that the pion-exchange terms considered here may constitute useful ingredients for constructing EDFs from first principles.
However, these conclusions are based on a certain optimization setting. 
We believe it would, therefore, be of interest to study if the improvement observed here also carries over to other optimization settings.

\section*{Acknowledgments}
I thank Jared O'Neal for helpful discussions and benchmarks of the covariance-matrix computations.
Calculations for this research were in part conducted on the Lichtenberg high performance computer of TU Darmstadt.

\bibliography{Literature}

\begin{thebibliography}{26}%
\makeatletter
\providecommand \@ifxundefined [1]{%
 \@ifx{#1\undefined}
}%
\providecommand \@ifnum [1]{%
 \ifnum #1\expandafter \@firstoftwo
 \else \expandafter \@secondoftwo
 \fi
}%
\providecommand \@ifx [1]{%
 \ifx #1\expandafter \@firstoftwo
 \else \expandafter \@secondoftwo
 \fi
}%
\providecommand \natexlab [1]{#1}%
\providecommand \enquote  [1]{``#1''}%
\providecommand \bibnamefont  [1]{#1}%
\providecommand \bibfnamefont [1]{#1}%
\providecommand \citenamefont [1]{#1}%
\providecommand \href@noop [0]{\@secondoftwo}%
\providecommand \href [0]{\begingroup \@sanitize@url \@href}%
\providecommand \@href[1]{\@@startlink{#1}\@@href}%
\providecommand \@@href[1]{\endgroup#1\@@endlink}%
\providecommand \@sanitize@url [0]{\catcode `\\12\catcode `\$12\catcode `\&12\catcode `\#12\catcode `\^12\catcode `\_12\catcode `\%12\relax}%
\providecommand \@@startlink[1]{}%
\providecommand \@@endlink[0]{}%
\providecommand \url  [0]{\begingroup\@sanitize@url \@url }%
\providecommand \@url [1]{\endgroup\@href {#1}{\urlprefix }}%
\providecommand \urlprefix  [0]{URL }%
\providecommand \Eprint [0]{\href }%
\providecommand \doibase [0]{https://doi.org/}%
\providecommand \selectlanguage [0]{\@gobble}%
\providecommand \bibinfo  [0]{\@secondoftwo}%
\providecommand \bibfield  [0]{\@secondoftwo}%
\providecommand \translation [1]{[#1]}%
\providecommand \BibitemOpen [0]{}%
\providecommand \bibitemStop [0]{}%
\providecommand \bibitemNoStop [0]{.\EOS\space}%
\providecommand \EOS [0]{\spacefactor3000\relax}%
\providecommand \BibitemShut  [1]{\csname bibitem#1\endcsname}%
\let\auto@bib@innerbib\@empty
\bibitem [{\citenamefont {Bender}\ \emph {et~al.}(2003)\citenamefont {Bender}, \citenamefont {Heenen},\ and\ \citenamefont {Reinhard}}]{Bend03RMP}%
  \BibitemOpen
  \bibfield  {author} {\bibinfo {author} {\bibfnamefont {M.}~\bibnamefont {Bender}}, \bibinfo {author} {\bibfnamefont {P.-H.}\ \bibnamefont {Heenen}},\ and\ \bibinfo {author} {\bibfnamefont {P.-G.}\ \bibnamefont {Reinhard}},\ }\href {https://doi.org/10.1103/RevModPhys.75.121} {\bibfield  {journal} {\bibinfo  {journal} {Rev. Mod. Phys.}\ }\textbf {\bibinfo {volume} {75}},\ \bibinfo {pages} {121} (\bibinfo {year} {2003})}\BibitemShut {NoStop}%
\bibitem [{\citenamefont {Schunck}(2019)}]{Schu19EDFBook}%
  \BibitemOpen
  \bibinfo {editor} {\bibfnamefont {N.}~\bibnamefont {Schunck}},\ ed.,\ \href {https://doi.org/10.1088/2053-2563/aae0ed} {\emph {\bibinfo {title} {Energy Density Functional Methods for Atomic Nuclei}}}\ (\bibinfo  {publisher} {IOP Publishing},\ \bibinfo {address} {Bristol},\ \bibinfo {year} {2019})\BibitemShut {NoStop}%
\bibitem [{\citenamefont {Mumpower}\ \emph {et~al.}(2016)\citenamefont {Mumpower}, \citenamefont {Surman}, \citenamefont {McLaughlin},\ and\ \citenamefont {Aprahamian}}]{Mump16IndiRPro}%
  \BibitemOpen
  \bibfield  {author} {\bibinfo {author} {\bibfnamefont {M.~R.}\ \bibnamefont {Mumpower}}, \bibinfo {author} {\bibfnamefont {R.}~\bibnamefont {Surman}}, \bibinfo {author} {\bibfnamefont {G.~C.}\ \bibnamefont {McLaughlin}},\ and\ \bibinfo {author} {\bibfnamefont {A.}~\bibnamefont {Aprahamian}},\ }\href {https://doi.org/10.1016/j.ppnp.2015.09.001} {\bibfield  {journal} {\bibinfo  {journal} {Prog. Part. Nucl. Phys.}\ }\textbf {\bibinfo {volume} {86}},\ \bibinfo {pages} {86} (\bibinfo {year} {2016})},\ \bibinfo {note} {corrigendum: \href{https://doi.org/10.1016/j.ppnp.2015.12.002}{Prog. Part. Nucl. Phys. \textbf{87}, 116 (2016)}},\ \Eprint {https://arxiv.org/abs/1508.07352} {arXiv:1508.07352 [nucl-th]} \BibitemShut {NoStop}%
\bibitem [{\citenamefont {McDonnell}\ \emph {et~al.}(2015)\citenamefont {McDonnell}, \citenamefont {Schunck}, \citenamefont {Higdon}, \citenamefont {Sarich}, \citenamefont {Wild},\ and\ \citenamefont {Nazarewicz}}]{McDo15UncQuaEDF}%
  \BibitemOpen
  \bibfield  {author} {\bibinfo {author} {\bibfnamefont {J.~D.}\ \bibnamefont {McDonnell}}, \bibinfo {author} {\bibfnamefont {N.}~\bibnamefont {Schunck}}, \bibinfo {author} {\bibfnamefont {D.}~\bibnamefont {Higdon}}, \bibinfo {author} {\bibfnamefont {J.}~\bibnamefont {Sarich}}, \bibinfo {author} {\bibfnamefont {S.~M.}\ \bibnamefont {Wild}},\ and\ \bibinfo {author} {\bibfnamefont {W.}~\bibnamefont {Nazarewicz}},\ }\href {https://doi.org/10.1103/PhysRevLett.114.122501} {\bibfield  {journal} {\bibinfo  {journal} {Phys. Rev. Lett.}\ }\textbf {\bibinfo {volume} {114}},\ \bibinfo {pages} {122501} (\bibinfo {year} {2015})},\ \Eprint {https://arxiv.org/abs/1501.03572} {arXiv:1501.03572 [nucl-th]} \BibitemShut {NoStop}%
\bibitem [{\citenamefont {Kortelainen}\ \emph {et~al.}(2008)\citenamefont {Kortelainen}, \citenamefont {Dobaczewski}, \citenamefont {Mizuyama},\ and\ \citenamefont {Toivanen}}]{Kort08SkyrmeSP}%
  \BibitemOpen
  \bibfield  {author} {\bibinfo {author} {\bibfnamefont {M.}~\bibnamefont {Kortelainen}}, \bibinfo {author} {\bibfnamefont {J.}~\bibnamefont {Dobaczewski}}, \bibinfo {author} {\bibfnamefont {K.}~\bibnamefont {Mizuyama}},\ and\ \bibinfo {author} {\bibfnamefont {J.}~\bibnamefont {Toivanen}},\ }\href {https://doi.org/10.1103/PhysRevC.77.064307} {\bibfield  {journal} {\bibinfo  {journal} {Phys. Rev. C}\ }\textbf {\bibinfo {volume} {77}},\ \bibinfo {pages} {064307} (\bibinfo {year} {2008})},\ \Eprint {https://arxiv.org/abs/0803.2291} {arXiv:0803.2291 [nucl-th]} \BibitemShut {NoStop}%
\bibitem [{\citenamefont {Duguet}\ \emph {et~al.}(2023)\citenamefont {Duguet}, \citenamefont {Ebran}, \citenamefont {Frosini}, \citenamefont {Hergert},\ and\ \citenamefont {Som\`a}}]{Dugu22PGCMEDF}%
  \BibitemOpen
  \bibfield  {author} {\bibinfo {author} {\bibfnamefont {T.}~\bibnamefont {Duguet}}, \bibinfo {author} {\bibfnamefont {J.-P.}\ \bibnamefont {Ebran}}, \bibinfo {author} {\bibfnamefont {M.}~\bibnamefont {Frosini}}, \bibinfo {author} {\bibfnamefont {H.}~\bibnamefont {Hergert}},\ and\ \bibinfo {author} {\bibfnamefont {V.}~\bibnamefont {Som\`a}},\ }\href {https://doi.org/10.1140/epja/s10050-023-00914-y} {\bibfield  {journal} {\bibinfo  {journal} {Eur. Phys. J. A}\ }\textbf {\bibinfo {volume} {59}},\ \bibinfo {pages} {13} (\bibinfo {year} {2023})},\ \Eprint {https://arxiv.org/abs/2209.03424} {arXiv:2209.03424 [nucl-th]} \BibitemShut {NoStop}%
\bibitem [{\citenamefont {Bulgac}\ \emph {et~al.}(2018)\citenamefont {Bulgac}, \citenamefont {Forbes}, \citenamefont {Jin}, \citenamefont {Navarro~Perez},\ and\ \citenamefont {Schunck}}]{Bulg18SeaLL1}%
  \BibitemOpen
  \bibfield  {author} {\bibinfo {author} {\bibfnamefont {A.}~\bibnamefont {Bulgac}}, \bibinfo {author} {\bibfnamefont {M.~M.}\ \bibnamefont {Forbes}}, \bibinfo {author} {\bibfnamefont {S.}~\bibnamefont {Jin}}, \bibinfo {author} {\bibfnamefont {R.}~\bibnamefont {Navarro~Perez}},\ and\ \bibinfo {author} {\bibfnamefont {N.}~\bibnamefont {Schunck}},\ }\href {https://doi.org/10.1103/PhysRevC.97.044313} {\bibfield  {journal} {\bibinfo  {journal} {Phys. Rev. C}\ }\textbf {\bibinfo {volume} {97}},\ \bibinfo {pages} {044313} (\bibinfo {year} {2018})},\ \Eprint {https://arxiv.org/abs/1708.08771} {arXiv:1708.08771 [nucl-th]} \BibitemShut {NoStop}%
\bibitem [{\citenamefont {Bonnard}\ \emph {et~al.}(2018)\citenamefont {Bonnard}, \citenamefont {Grasso},\ and\ \citenamefont {Lacroix}}]{Bonn18EDFNDrop}%
  \BibitemOpen
  \bibfield  {author} {\bibinfo {author} {\bibfnamefont {J.}~\bibnamefont {Bonnard}}, \bibinfo {author} {\bibfnamefont {M.}~\bibnamefont {Grasso}},\ and\ \bibinfo {author} {\bibfnamefont {D.}~\bibnamefont {Lacroix}},\ }\href {https://doi.org/10.1103/PhysRevC.98.034319} {\bibfield  {journal} {\bibinfo  {journal} {Phys. Rev. C}\ }\textbf {\bibinfo {volume} {98}},\ \bibinfo {pages} {034319} (\bibinfo {year} {2018})},\ \bibinfo {note} {erratum: \href{https://doi.org/10.1103/PhysRevC.103.039901}{Phys. Rev. C \textbf{103}, 039901(E) (2021)}},\ \Eprint {https://arxiv.org/abs/1806.01084} {arXiv:1806.01084 [nucl-th]} \BibitemShut {NoStop}%
\bibitem [{\citenamefont {Furnstahl}(2020)}]{Furn20EDFasEFT}%
  \BibitemOpen
  \bibfield  {author} {\bibinfo {author} {\bibfnamefont {R.~J.}\ \bibnamefont {Furnstahl}},\ }\href {https://doi.org/10.1140/epja/s10050-020-00095-y} {\bibfield  {journal} {\bibinfo  {journal} {Eur. Phys. J. A}\ }\textbf {\bibinfo {volume} {56}},\ \bibinfo {pages} {85} (\bibinfo {year} {2020})},\ \Eprint {https://arxiv.org/abs/1906.00833} {arXiv:1906.00833 [nucl-th]} \BibitemShut {NoStop}%
\bibitem [{\citenamefont {Drut}\ \emph {et~al.}(2010)\citenamefont {Drut}, \citenamefont {Furnstahl},\ and\ \citenamefont {Platter}}]{Drut10PPNP}%
  \BibitemOpen
  \bibfield  {author} {\bibinfo {author} {\bibfnamefont {J.~E.}\ \bibnamefont {Drut}}, \bibinfo {author} {\bibfnamefont {R.~J.}\ \bibnamefont {Furnstahl}},\ and\ \bibinfo {author} {\bibfnamefont {L.}~\bibnamefont {Platter}},\ }\href {https://doi.org/10.1016/j.ppnp.2009.09.001} {\bibfield  {journal} {\bibinfo  {journal} {Prog. Part. Nucl. Phys.}\ }\textbf {\bibinfo {volume} {64}},\ \bibinfo {pages} {120} (\bibinfo {year} {2010})},\ \Eprint {https://arxiv.org/abs/0906.1463} {arXiv:0906.1463 [nucl-th]} \BibitemShut {NoStop}%
\bibitem [{\citenamefont {Gebremariam}\ \emph {et~al.}(2010)\citenamefont {Gebremariam}, \citenamefont {Duguet},\ and\ \citenamefont {Bogner}}]{Gebr10DME}%
  \BibitemOpen
  \bibfield  {author} {\bibinfo {author} {\bibfnamefont {B.}~\bibnamefont {Gebremariam}}, \bibinfo {author} {\bibfnamefont {T.}~\bibnamefont {Duguet}},\ and\ \bibinfo {author} {\bibfnamefont {S.~K.}\ \bibnamefont {Bogner}},\ }\href {https://doi.org/10.1103/PhysRevC.82.014305} {\bibfield  {journal} {\bibinfo  {journal} {Phys. Rev. C}\ }\textbf {\bibinfo {volume} {82}},\ \bibinfo {pages} {014305} (\bibinfo {year} {2010})},\ \Eprint {https://arxiv.org/abs/0910.4979} {arXiv:0910.4979 [nucl-th]} \BibitemShut {NoStop}%
\bibitem [{\citenamefont {Zurek}\ \emph {et~al.}(2024)\citenamefont {Zurek}, \citenamefont {Bogner}, \citenamefont {Furnstahl}, \citenamefont {Navarro~P\'erez}, \citenamefont {Schunck},\ and\ \citenamefont {Schwenk}}]{Zurek24GUDE}%
  \BibitemOpen
  \bibfield  {author} {\bibinfo {author} {\bibfnamefont {L.}~\bibnamefont {Zurek}}, \bibinfo {author} {\bibfnamefont {S.~K.}\ \bibnamefont {Bogner}}, \bibinfo {author} {\bibfnamefont {R.~J.}\ \bibnamefont {Furnstahl}}, \bibinfo {author} {\bibfnamefont {R.}~\bibnamefont {Navarro~P\'erez}}, \bibinfo {author} {\bibfnamefont {N.}~\bibnamefont {Schunck}},\ and\ \bibinfo {author} {\bibfnamefont {A.}~\bibnamefont {Schwenk}},\ }\href {https://doi.org/10.1103/PhysRevC.109.014319} {\bibfield  {journal} {\bibinfo  {journal} {Phys. Rev. C}\ }\textbf {\bibinfo {volume} {109}},\ \bibinfo {pages} {014319} (\bibinfo {year} {2024})},\ \Eprint {https://arxiv.org/abs/2307.13568} {arXiv:2307.13568 [nucl-th]} \BibitemShut {NoStop}%
\bibitem [{\citenamefont {Sun}\ \emph {et~al.}(2022)\citenamefont {Sun}, \citenamefont {Bell}, \citenamefont {Hagen},\ and\ \citenamefont {Papenbrock}}]{Sun22RenormCC}%
  \BibitemOpen
  \bibfield  {author} {\bibinfo {author} {\bibfnamefont {Z.~H.}\ \bibnamefont {Sun}}, \bibinfo {author} {\bibfnamefont {C.~A.}\ \bibnamefont {Bell}}, \bibinfo {author} {\bibfnamefont {G.}~\bibnamefont {Hagen}},\ and\ \bibinfo {author} {\bibfnamefont {T.}~\bibnamefont {Papenbrock}},\ }\href {https://doi.org/10.1103/PhysRevC.106.L061302} {\bibfield  {journal} {\bibinfo  {journal} {Phys. Rev. C}\ }\textbf {\bibinfo {volume} {106}},\ \bibinfo {pages} {L061302} (\bibinfo {year} {2022})},\ \Eprint {https://arxiv.org/abs/2205.12990} {arXiv:2205.12990 [nucl-th]} \BibitemShut {NoStop}%
\bibitem [{\citenamefont {Krebs}\ \emph {et~al.}(2007)\citenamefont {Krebs}, \citenamefont {Epelbaum},\ and\ \citenamefont {Mei{\ss}ner}}]{Kreb07Deltas}%
  \BibitemOpen
  \bibfield  {author} {\bibinfo {author} {\bibfnamefont {H.}~\bibnamefont {Krebs}}, \bibinfo {author} {\bibfnamefont {E.}~\bibnamefont {Epelbaum}},\ and\ \bibinfo {author} {\bibfnamefont {U.-G.}\ \bibnamefont {Mei{\ss}ner}},\ }\href {https://doi.org/10.1140/epja/i2007-10372-y} {\bibfield  {journal} {\bibinfo  {journal} {Eur. Phys. J. A}\ }\textbf {\bibinfo {volume} {32}},\ \bibinfo {pages} {127} (\bibinfo {year} {2007})},\ \Eprint {https://arxiv.org/abs/nucl-th/0703087} {arXiv:nucl-th/0703087} \BibitemShut {NoStop}%
\bibitem [{\citenamefont {Kortelainen}\ \emph {et~al.}(2014)\citenamefont {Kortelainen}, \citenamefont {McDonnell}, \citenamefont {Nazarewicz}, \citenamefont {Olsen}, \citenamefont {Reinhard}, \citenamefont {Sarich}, \citenamefont {Schunck}, \citenamefont {Wild}, \citenamefont {Davesne}, \citenamefont {Erler},\ and\ \citenamefont {Pastore}}]{Kort14UNEDF2}%
  \BibitemOpen
  \bibfield  {author} {\bibinfo {author} {\bibfnamefont {M.}~\bibnamefont {Kortelainen}}, \bibinfo {author} {\bibfnamefont {J.}~\bibnamefont {McDonnell}}, \bibinfo {author} {\bibfnamefont {W.}~\bibnamefont {Nazarewicz}}, \bibinfo {author} {\bibfnamefont {E.}~\bibnamefont {Olsen}}, \bibinfo {author} {\bibfnamefont {P.-G.}\ \bibnamefont {Reinhard}}, \bibinfo {author} {\bibfnamefont {J.}~\bibnamefont {Sarich}}, \bibinfo {author} {\bibfnamefont {N.}~\bibnamefont {Schunck}}, \bibinfo {author} {\bibfnamefont {S.~M.}\ \bibnamefont {Wild}}, \bibinfo {author} {\bibfnamefont {D.}~\bibnamefont {Davesne}}, \bibinfo {author} {\bibfnamefont {J.}~\bibnamefont {Erler}},\ and\ \bibinfo {author} {\bibfnamefont {A.}~\bibnamefont {Pastore}},\ }\href {https://doi.org/10.1103/PhysRevC.89.054314} {\bibfield  {journal} {\bibinfo  {journal} {Phys. Rev. C}\ }\textbf {\bibinfo {volume} {89}},\ \bibinfo {pages} {054314} (\bibinfo {year} {2014})},\ \Eprint {https://arxiv.org/abs/1312.1746} {arXiv:1312.1746 [nucl-th]} \BibitemShut
  {NoStop}%
\bibitem [{\citenamefont {Navarro~Pérez}\ \emph {et~al.}(2018)\citenamefont {Navarro~Pérez}, \citenamefont {Schunck}, \citenamefont {Dyhdalo}, \citenamefont {Furnstahl},\ and\ \citenamefont {Bogner}}]{Nava18DMEEDF}%
  \BibitemOpen
  \bibfield  {author} {\bibinfo {author} {\bibfnamefont {R.}~\bibnamefont {Navarro~Pérez}}, \bibinfo {author} {\bibfnamefont {N.}~\bibnamefont {Schunck}}, \bibinfo {author} {\bibfnamefont {A.}~\bibnamefont {Dyhdalo}}, \bibinfo {author} {\bibfnamefont {R.~J.}\ \bibnamefont {Furnstahl}},\ and\ \bibinfo {author} {\bibfnamefont {S.~K.}\ \bibnamefont {Bogner}},\ }\href {https://doi.org/10.1103/PhysRevC.97.054304} {\bibfield  {journal} {\bibinfo  {journal} {Phys. Rev. C}\ }\textbf {\bibinfo {volume} {97}},\ \bibinfo {pages} {054304} (\bibinfo {year} {2018})},\ \Eprint {https://arxiv.org/abs/1801.08615} {arXiv:1801.08615 [nucl-th]} \BibitemShut {NoStop}%
\bibitem [{\citenamefont {Schunck}\ \emph {et~al.}(2020)\citenamefont {Schunck}, \citenamefont {O'Neal}, \citenamefont {Grosskopf}, \citenamefont {Lawrence},\ and\ \citenamefont {Wild}}]{Schu20EDFCalibr}%
  \BibitemOpen
  \bibfield  {author} {\bibinfo {author} {\bibfnamefont {N.}~\bibnamefont {Schunck}}, \bibinfo {author} {\bibfnamefont {J.}~\bibnamefont {O'Neal}}, \bibinfo {author} {\bibfnamefont {M.}~\bibnamefont {Grosskopf}}, \bibinfo {author} {\bibfnamefont {E.}~\bibnamefont {Lawrence}},\ and\ \bibinfo {author} {\bibfnamefont {S.~M.}\ \bibnamefont {Wild}},\ }\href {https://doi.org/10.1088/1361-6471/ab8745} {\bibfield  {journal} {\bibinfo  {journal} {J. Phys. G}\ }\textbf {\bibinfo {volume} {47}},\ \bibinfo {pages} {074001} (\bibinfo {year} {2020})},\ \Eprint {https://arxiv.org/abs/2003.12207} {arXiv:2003.12207 [nucl-th]} \BibitemShut {NoStop}%
\bibitem [{\citenamefont {Donaldson}\ and\ \citenamefont {Schnabel}(1987)}]{Donaldson87NonlinearLS}%
  \BibitemOpen
  \bibfield  {author} {\bibinfo {author} {\bibfnamefont {J.~R.}\ \bibnamefont {Donaldson}}\ and\ \bibinfo {author} {\bibfnamefont {R.~B.}\ \bibnamefont {Schnabel}},\ }\href {https://doi.org/10.1080/00401706.1987.10488184} {\bibfield  {journal} {\bibinfo  {journal} {Technometrics}\ }\textbf {\bibinfo {volume} {29}},\ \bibinfo {pages} {67} (\bibinfo {year} {1987})}\BibitemShut {NoStop}%
\bibitem [{\citenamefont {Seber}\ and\ \citenamefont {Wild}(1989)}]{Seber2003NonlinearRegressionBook}%
  \BibitemOpen
  \bibfield  {author} {\bibinfo {author} {\bibfnamefont {G.~A.~F.}\ \bibnamefont {Seber}}\ and\ \bibinfo {author} {\bibfnamefont {C.~J.}\ \bibnamefont {Wild}},\ }\href {https://doi.org/https://doi.org/10.1002/0471725315} {\emph {\bibinfo {title} {Nonlinear Regression}}}\ (\bibinfo  {publisher} {John Wiley \& Sons, Ltd},\ \bibinfo {year} {1989})\BibitemShut {NoStop}%
\bibitem [{\citenamefont {Reinhard}\ \emph {et~al.}(2024)\citenamefont {Reinhard}, \citenamefont {O'Neal}, \citenamefont {Wild},\ and\ \citenamefont {Nazarewicz}}]{Reinhard24Fayans}%
  \BibitemOpen
  \bibfield  {author} {\bibinfo {author} {\bibfnamefont {P.-G.}\ \bibnamefont {Reinhard}}, \bibinfo {author} {\bibfnamefont {J.}~\bibnamefont {O'Neal}}, \bibinfo {author} {\bibfnamefont {S.~M.}\ \bibnamefont {Wild}},\ and\ \bibinfo {author} {\bibfnamefont {W.}~\bibnamefont {Nazarewicz}},\ }\href {https://doi.org/10.1088/1361-6471/ad633a} {\bibfield  {journal} {\bibinfo  {journal} {J. Phys. G}\ }\textbf {\bibinfo {volume} {51}},\ \bibinfo {pages} {105101} (\bibinfo {year} {2024})},\ \Eprint {https://arxiv.org/abs/2402.15380} {arXiv:2402.15380 [nucl-th]} \BibitemShut {NoStop}%
\bibitem [{\citenamefont {Kortelainen}\ \emph {et~al.}(2010)\citenamefont {Kortelainen}, \citenamefont {Lesinski}, \citenamefont {Mor{\'e}}, \citenamefont {Nazarewicz}, \citenamefont {Sarich}, \citenamefont {Schunck}, \citenamefont {Stoitsov},\ and\ \citenamefont {Wild}}]{Kort10edf}%
  \BibitemOpen
  \bibfield  {author} {\bibinfo {author} {\bibfnamefont {M.}~\bibnamefont {Kortelainen}}, \bibinfo {author} {\bibfnamefont {T.}~\bibnamefont {Lesinski}}, \bibinfo {author} {\bibfnamefont {J.}~\bibnamefont {Mor{\'e}}}, \bibinfo {author} {\bibfnamefont {W.}~\bibnamefont {Nazarewicz}}, \bibinfo {author} {\bibfnamefont {J.}~\bibnamefont {Sarich}}, \bibinfo {author} {\bibfnamefont {N.}~\bibnamefont {Schunck}}, \bibinfo {author} {\bibfnamefont {M.~V.}\ \bibnamefont {Stoitsov}},\ and\ \bibinfo {author} {\bibfnamefont {S.}~\bibnamefont {Wild}},\ }\href {https://doi.org/10.1103/PhysRevC.82.024313} {\bibfield  {journal} {\bibinfo  {journal} {Phys. Rev. C}\ }\textbf {\bibinfo {volume} {82}},\ \bibinfo {pages} {024313} (\bibinfo {year} {2010})},\ \Eprint {https://arxiv.org/abs/1005.5145} {arXiv:1005.5145 [nucl-th]} \BibitemShut {NoStop}%
\bibitem [{\citenamefont {Wild}\ \emph {et~al.}(2015)\citenamefont {Wild}, \citenamefont {Sarich},\ and\ \citenamefont {Schunck}}]{Wild15POUNDEDF}%
  \BibitemOpen
  \bibfield  {author} {\bibinfo {author} {\bibfnamefont {S.~M.}\ \bibnamefont {Wild}}, \bibinfo {author} {\bibfnamefont {J.}~\bibnamefont {Sarich}},\ and\ \bibinfo {author} {\bibfnamefont {N.}~\bibnamefont {Schunck}},\ }\href {https://doi.org/10.1088/0954-3899/42/3/034031} {\bibfield  {journal} {\bibinfo  {journal} {J. Phys. G}\ }\textbf {\bibinfo {volume} {42}},\ \bibinfo {pages} {034031} (\bibinfo {year} {2015})},\ \Eprint {https://arxiv.org/abs/1406.5464} {arXiv:1406.5464 [physics.comp-ph]} \BibitemShut {NoStop}%
\bibitem [{\citenamefont {Nazarewicz}\ \emph {et~al.}(2014)\citenamefont {Nazarewicz}, \citenamefont {Reinhard}, \citenamefont {Satula},\ and\ \citenamefont {Vretenar}}]{Naza14SymEnDFT}%
  \BibitemOpen
  \bibfield  {author} {\bibinfo {author} {\bibfnamefont {W.}~\bibnamefont {Nazarewicz}}, \bibinfo {author} {\bibfnamefont {P.-G.}\ \bibnamefont {Reinhard}}, \bibinfo {author} {\bibfnamefont {W.}~\bibnamefont {Satula}},\ and\ \bibinfo {author} {\bibfnamefont {D.}~\bibnamefont {Vretenar}},\ }\href {https://doi.org/10.1140/epja/i2014-14020-3} {\bibfield  {journal} {\bibinfo  {journal} {Eur. Phys. J. A}\ }\textbf {\bibinfo {volume} {50}},\ \bibinfo {pages} {20} (\bibinfo {year} {2014})},\ \Eprint {https://arxiv.org/abs/1307.5782} {arXiv:1307.5782 [nucl-th]} \BibitemShut {NoStop}%
\bibitem [{\citenamefont {Dobaczewski}\ \emph {et~al.}(2014)\citenamefont {Dobaczewski}, \citenamefont {Nazarewicz},\ and\ \citenamefont {Reinhard}}]{Doba14error}%
  \BibitemOpen
  \bibfield  {author} {\bibinfo {author} {\bibfnamefont {J.}~\bibnamefont {Dobaczewski}}, \bibinfo {author} {\bibfnamefont {W.}~\bibnamefont {Nazarewicz}},\ and\ \bibinfo {author} {\bibfnamefont {P.~G.}\ \bibnamefont {Reinhard}},\ }\href {https://doi.org/10.1088/0954-3899/41/7/074001} {\bibfield  {journal} {\bibinfo  {journal} {J. Phys. G}\ }\textbf {\bibinfo {volume} {41}},\ \bibinfo {pages} {074001} (\bibinfo {year} {2014})},\ \Eprint {https://arxiv.org/abs/1402.4657} {arXiv:1402.4657 [nucl-th]} \BibitemShut {NoStop}%
\bibitem [{\citenamefont {Wang}\ \emph {et~al.}(2021)\citenamefont {Wang}, \citenamefont {Huang}, \citenamefont {Kondev}, \citenamefont {Audi},\ and\ \citenamefont {Naimi}}]{Wang21AME20}%
  \BibitemOpen
  \bibfield  {author} {\bibinfo {author} {\bibfnamefont {M.}~\bibnamefont {Wang}}, \bibinfo {author} {\bibfnamefont {W.~J.}\ \bibnamefont {Huang}}, \bibinfo {author} {\bibfnamefont {F.~G.}\ \bibnamefont {Kondev}}, \bibinfo {author} {\bibfnamefont {G.}~\bibnamefont {Audi}},\ and\ \bibinfo {author} {\bibfnamefont {S.}~\bibnamefont {Naimi}},\ }\href {https://doi.org/10.1088/1674-1137/abddaf} {\bibfield  {journal} {\bibinfo  {journal} {Chin. Phys. C}\ }\textbf {\bibinfo {volume} {45}},\ \bibinfo {pages} {030003} (\bibinfo {year} {2021})}\BibitemShut {NoStop}%
\bibitem [{\citenamefont {Angeli}\ and\ \citenamefont {Marinova}(2013)}]{Ange13rch}%
  \BibitemOpen
  \bibfield  {author} {\bibinfo {author} {\bibfnamefont {I.}~\bibnamefont {Angeli}}\ and\ \bibinfo {author} {\bibfnamefont {K.}~\bibnamefont {Marinova}},\ }\href {https://doi.org/10.1016/j.adt.2011.12.006} {\bibfield  {journal} {\bibinfo  {journal} {Atom. Data Nucl. Data Tabl.}\ }\textbf {\bibinfo {volume} {99}},\ \bibinfo {pages} {69} (\bibinfo {year} {2013})}\BibitemShut {NoStop}%
\end{thebibliography}%

\end{document}